\documentstyle[aps,twocolumn,epsf,psfig]{revtex}

\begin{document}

\bibliographystyle{prsty}
\draft

\title {Electronic Kinetic Energy in the Polaron Ground State and the Location of the Self-Trapping Transition}

\author{A.~H.~Romero${^{1,3}}$, David W. Brown${^2}$ and Katja Lindenberg${^3}$}

\address
{${^1}$
Department of Physics,\\
University of California, San Diego, La Jolla, CA 92093-0354}

\address
{${^2}$
Institute for Nonlinear Science,\\
University of California, San Diego, La Jolla, CA 92093-0402}

\address
{${^3}$
Department of Chemistry and Biochemistry,\\
University of California, San Diego, La Jolla, CA 92093-0340} 

\date{\today} 

\maketitle

\begin{abstract}

In this paper we discuss the electronic kinetic energy in the
ground state of the Holstein Hamiltonian in one space
dimension using the Global-Local variational method
together with perturbation theory at weak and strong coupling.
The electronic kinetic energy is detailed over a broad region of the
polaron parameter space, and is analyzed to accurately determine the
location of the self-trapping transition.
The self-trapping line so determined separates the small polaron
regime from the large polaron regime, constituting a polaron
phase diagram.

\end{abstract}
\vspace{0.5in}


\narrowtext

\section{Introduction}

Despite the vast amount of effort and creativity that has been applied to the polaron problem for more than half a century, there remain significant aspects of polaron structure and behavior that have defied satisfying explanation and quantitative description.
Central among these is the so-called self-trapping transition.
The self-trapping transition describes the change in polaron structure from being large-polaron-like over one region of the polaron parameter space to being small-polaron-like in another.
Perturbation theories exist providing asymptotically accurate descriptions of polaron structure and properties on each side of this transition; however, the intermediate-coupling region in which the transition lies has proven very resistant to analysis.

Our starting point is the 1-D Holstein Hamiltonian ~\cite{Holstein59a,Holstein59b}
\begin{eqnarray}
\hat{H} &=& 
- J \sum_n a_n^{\dagger} ( a_{n+1} + a_{n-1} ) +
\hbar \omega \sum_n b_n^{\dagger} b_n \nonumber \\
& & - g \hbar \omega \sum_n a_n^{\dagger} a_n ( b_n^{\dagger} + b_n ) ~,
\end{eqnarray}
in which $a_n^\dagger$ creates a single electronic excitation in the rigid-lattice Wannier state at site $n$, and $b_n^\dagger$ creates a quantum of vibrational energy  $\hbar \omega$ in the Einstein oscillator at site $n$. 
The hopping matrix element connecting nearest neighbors is given by $J$, and $g$ is the electron-phonon coupling  strength.

The last few years have seen considerable progress in the application of polaron-friendly theoretical techniques to the Holstein model; these include
variational techniques ~\cite{Brown97a,Zhao97a,Zhao97b,Romero97},
cluster diagonalization ~\cite{Capone97b,Wellein97a,deMello97,Alexandrov94a}
dynamical mean field theory ~\cite{Ciuchi97,Ciuchi95},
perturbation theory ~\cite{Gogolin82,Capone97a,Marsiglio95,Stephan96},
density matrix renormalization group (DMRG) ~\cite{Jeckelmann97,Jeckelmann98} and
quantum Monte Carlo simulations (QMC) ~\cite{McKenzie96,Lagendijk85,DeRaedt84,DeRaedt83,DeRaedt97}.
While each methodology has its own merits, each also has its weaknesses and only few of them possess the scope to be accurate over broad regions of the parameter space, especially at the intermediate electron-phonon coupling well away from all limits.

We approach the problem variationally using the Global-Local method, introduced by Brown {\it et al.} \cite{Brown97a} for the determination of polaron Bloch states and band energies across the entire Brillouin zone ($\kappa \in \{ - \pi , + \pi \}$).
In this paper, we shall focus on the polaron {\it ground state} ($\kappa = 0$), and in particular on the kinetic energy in the ground state.
Details regarding the variational method and its numerical implementation can be found in ref~\cite{Brown97a}.
The numerical data presented herein were computed on a 1-D lattice of 32 sites with periodic boundary conditions.

\section{The Kinetic Energy}

We determine the kinetic energy variationally, and compare our results with both weak-coupling perturbation theory and strong-coupling perturbation theory, as well as selected results of other methods in the intermediate regime.
In each case, the kinetic energy is computed as the expectation value
\begin{equation}
E_{kin} = \langle \psi | - J \sum_n a_n^{\dagger} ( a_{n+1} + a_{n-1} ) | \psi \rangle
\end{equation}
in the appropriate ground state $| \psi \rangle$.

\begin{figure}[htb]
\begin{center}
\leavevmode
\epsfxsize = 3.7in
\epsffile{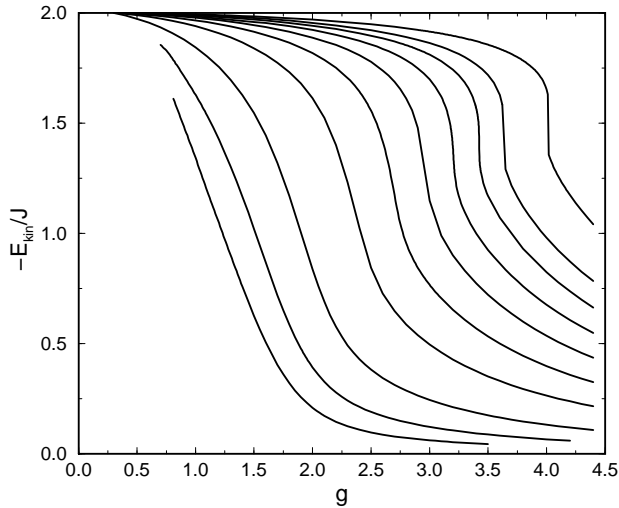}
\end{center}
\caption
{
The polaron kinetic energy $E_{kin}$ as a function of the electron-phonon coupling.
Curves left to right: $J/\hbar \omega = 0.25$, $ 0.5$, $ 1.0$, $ 2.0$, $ 3.0$, $ 4.0$, $ 5.0$, $ 6.0$, $ 7.0 $ and $9.0$. 
}
\label{fig:energies}
\end{figure}

Our numerical results are summarized in Figure~\ref{fig:energies}.
Each curve in Figure~\ref{fig:energies} contains 80-200 data points, with a higher density of points clustered in the steeper portions of each curve where the self-trapping transition is expected to be found.
No smoothing has been performed; each "curve" is a polygonal arc connecting computed energies.
Over most of the parameter space we have investigated, computational errors are smaller than can be meaningfully conveyed in with any resolvable symbol; the principal exception is at weak coupling and smaller $J$ values where decreasing sensitivity of the ground state energy to certain details of polaron structure eventually hampers convergence.

With increasing adiabaticity, here beginning at $J/\hbar \omega \approx 7$, the ability of the variational method to represent the complexity of polaron structure in the immediate vicinity of the self-trapping transition eventually is overtaxed, and discontinuities appear in estimated quantities such as the kinetic energy.
Although the {\it value} of the kinetic energy in the immediate vicinity of such anomalies is necessarily distorted, the location of the discontinuities continues to provide reasonable estimates for the location of the self-trapping transition, and outside of a narrow region, quantitative accuracy remains excellent. \cite{Romero97}
For such reasons, we retain in the following data for the cases $J/\hbar \omega = 7$ and $9$.

The weak coupling regime can be understood through comparison with weak-coupling perturbation theory, which to second order in the coupling constant yields
\begin{equation}
E_{kin}^{WC} \sim  - 2 J + \frac{2J g^2}{(1 + 4J/\hbar \omega )^{3/2}}
\label{eq:weak}
\end{equation}
Characteristically, the kinetic energy is a weak function of the electron-phonon coupling below the self-trapping transition, and grows increasingly weak with increasing adiabaticity.
Owing to the minimal involvement of phonons in the polaron in this regime, the quasi-particle can be fairly characterized as a quasi-free electron with a slightly reduced bandwidth. 
It is characteristic of weak-coupling perturbation theory, however, that the integration of phonons into polaron structure is {\it under}represented, causing the electron to appear more "free" than it actually is.

The strong coupling regime can be understood through comparison with strong-coupling perturbation theory, which to second order in $J$ (following the Lang-Firsov transformation \cite{Lang63}) yields \cite{Gogolin82,Alexandrov95}
\begin{eqnarray}
E_{kin}^{LF} &\sim& -2 J e^{-g^2} - \frac {4J^2} {\hbar \omega} e^{-2g^2} [ f(2g^2) + f(g^2)]
\label{eq:strong}
 \\
&\sim& - 2  \frac{J^2}{g^2 \hbar \omega} ~~~ g \gg 1
\label{eq:stronger}
\end{eqnarray}
wherein $f(y) = {\rm Ei}(y) - \gamma - {\rm ln}(y)$, and ${\rm Ei}(y)$ is the exponential integral.
At very strong coupling, the kinetic energy decays to zero, suggesting that the dressed electron becomes essentially immobile relative to the quasi-free electron; significant, however, is the fact that this decay is not ultimately exponential in the coupling constant as results from the small polaron approximation (first order Lang-Firsov), but a much weaker inverse power as suggested by perturbative corrections (second-order Lang Firsov) \cite{Alexandrov98}.

A comparison of our numerical kinetic energies with weak (\ref{eq:weak}) and strong (\ref{eq:strong}) coupling perturbation theories is shown in Figure 2.
\begin{figure}[htb]
\begin{center}
\leavevmode
\epsfxsize = 3.5in
\epsffile{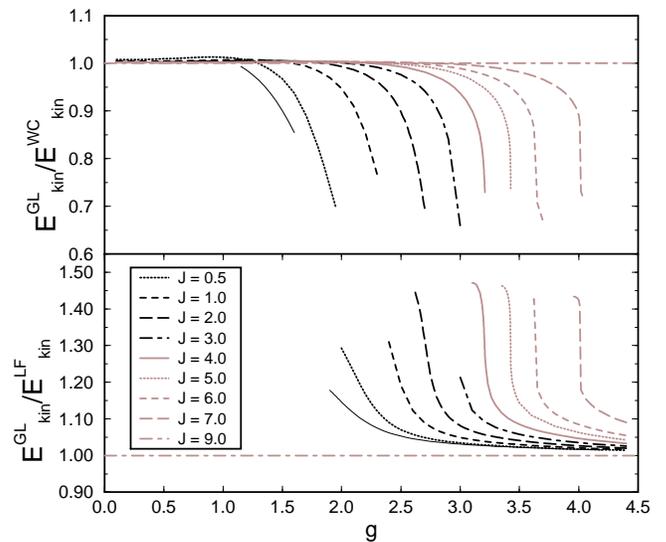}
\end{center}
\caption
{
Comparison of Global-Local kinetic energy, $E_{kin}^{GL}$, with weak-coupling perturbation theory, $E_{kin}^{WC}$ according to (\ref{eq:weak}) (upper panel), and with the Lang-Firsov approximation, $E_{kin}^{LF}$ according to (\ref{eq:strong}) (lower panel), as functions of the electron-phonon coupling. 
}
\label{fig:langfirsov}
\end{figure} 
For weak electron-phonon coupling, the agreement between our variational and perturbative kinetic energy is very good up to $g$ values quite close to the self-trapping transition.
For this weak coupling case, perturbation theory {\it over}estimates the kinetic energy value (i.e. phonon involvement in the polaron is greater than is captured by perturbation theory).

For strong coupling, our variational calculation approaches the second-order Lang-Firsov result above the self-trapping transition; however, deviations persist significantly into the strong coupling regime, with larger $J$'s converging to the Lang-Firsov result more slowly than smaller $J$'s.
Strong coupling perturbation theory systematically {\it under}estimates the kinetic energy (i.e. phonon involvement in the polaron is weaker than is embodied in Lang-Firsov). 

\begin{figure}[htb]
\begin{center}
\leavevmode
\epsfxsize = 3.75in
\epsffile{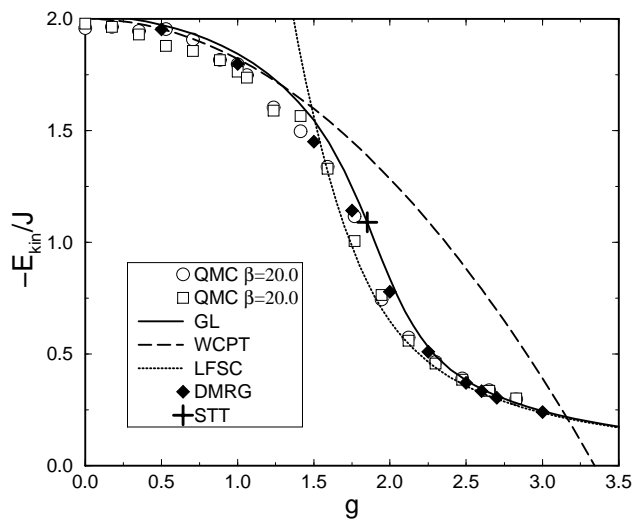}
\end{center}
\caption
{
Comparison of the kinetic energy $E_{kin}$ as determined by the Global-Local method (solid line) with weak coupling perturbation theory (dashed line), strong coupling perturbation theory (dotted line), quantum Monte Carlo simulation (scatter plot), and density matrix renormalization group (diamonds) for $J/\hbar \omega =1$.
The (+) symbol indicates the location of the self-trapping transition as determined using (\ref{eq:inflect}) on the GL data.
QMC data consists of two statistically-independent scans at the inverse temperature $\beta = 20.0$; data kindly provided by H. De Raedt \protect \cite{DeRaedt97} \protect.
DMRG data kindly provided by E. Jeckelmann \protect \cite{Jeckelmann98} \protect.
}
\label{fig:qmc}
\end{figure}

\section{The Self-Trapping Line}

The self-trapping transition is the more-or-less rapid change in polaron structure from that typical of large polarons (below the transition) to that typical of small polarons (above) as $g$ or $J$ are varied and is typically resolved as a feature that grows increasingly sharp with increasing adiabaticity.
The self-trapping transition is certainly the most exotic feature of the polaron phase diagram, and is intimately involved with, if not always ultimately responsible for, many of the difficulties encountered in polaron theory.
The {\it physical} transition is smooth at finite $g$ and $J/\hbar \omega$ \cite{Lowen88}; however, it is common for approximate descriptions of the phenomenon either to miss the transition completely, or to break down in some respect in its vicinity.

\begin{figure}[htb]
\begin{center}
\leavevmode
\epsfxsize = 3.75in
\epsffile{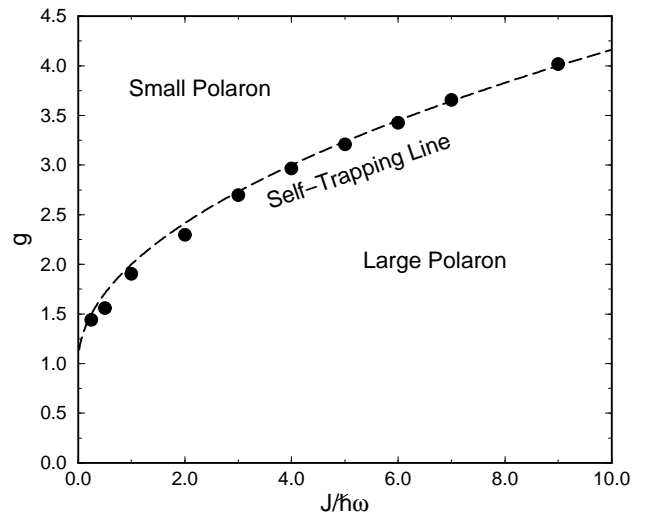}
\end{center}
\caption
{
Polaron phase diagram, showing the location of the self trapping transition.
Dots ($\bullet$) corresponds to our numerical calculation,
dashed line corresponds to the empirical curve $g_{ST} = 1 + \sqrt{J/\hbar \omega}$.
}
\label{fig:phasespace}
\end{figure}

Here, we will use the kinetic energy to define a transition on the physical rationale that this energy should perhaps most directly reflect the heavily-dressed and essentially immobile character of the self-trapped state ~\cite{DeRaedt84}.
As our objective criterion for locating this transition, we associate the transition with a particular coupling strength, $g_{ST}(J/\hbar \omega)$, where the kinetic energy changes {\it most rapidly} with respect to $g$ at fixed $J/\hbar \omega$; this point is identified by the extremum
\begin{equation}
\left. \frac{\partial ^2E_{kin}}{\partial  g^2} \right|_{g \equiv g_{ST}} = 0\;\;\;.
\label{eq:inflect}
\end{equation}
Since the sigmoidal shape of the kinetic energy is well resolved for all hopping integral values used in this work, a clear definition of the transition can be obtained from this criterion.
A summary of our calculation is shown in Figure~\ref{fig:phasespace}.

The function
\begin{equation}
g_{ST} = 1 + \sqrt{J/\hbar \omega}
\label{eq:gst}
\end{equation}
describes a strictly empirical curve that we have introduced to simply, conveniently, and accurately characterize the self-trapping line over the investigated regime.
No fit has been performed to obtain this curve.

The accuracy with which $g_{ST}$ describes the apparent self-trapping line far exceeds any other estimation currently available.
For example, one frequently finds the self-trapping transition associated with the apparent breakdown point of strong-coupling perturbation theory (at $\lambda \equiv g^2 \hbar \omega /2J \approx \frac 1 2$) or with the crossing points of weak and strong-coupling perturbation theory (see, e.g., Figure~\ref{fig:qmc}; neither of which compares as favorably as (\ref{eq:gst}) over the investigated regime.
There is reason to expect (\ref{eq:gst}) to continue to accurately locate the self-trapping transition {\it beyond} the investigated regime, since it is known from strong-coupling perturbation theory that the transition is associated with the limit \cite{Alexandrov95}
\begin{equation}
\lim_{g,J \rightarrow \infty} \lambda_{ST} = \lambda_c = \frac 1 2
\end{equation}
which is satisfied by (\ref{eq:gst}).

Although the numerical data in Figure~\ref{fig:phasespace} were generated from an analysis of one particular quantity, the kinetic energy, the same curve $g_{ST}$ can be expected to describe the appearance of the self-trapping transition in other physical quantities, verifying (\ref{eq:gst}) to be a robust as well as accurate locator of the self-trapping transition, as will be demonstrated elsewhere \cite{Romero98}.

\section*{acknowledgment}

This work was supported in part by the U.S. Department of Energy under Grant No.
 DE-FG03-86ER13606.
The authors gratefully acknowledge the cooperation of H. De Raedt and E. Jeckelmann for providing numerical values of data used in parts of this paper.

\bibliography{theory,books,experiment,temporal}

\end{document}